\documentclass{osa-article}
\journal{osajournal}
\articletype{Research Article}

\begin{document}

\title{Entanglement witness and entropy uncertainty of open quantum system under Zeno effect}

\author{Rongfang Liu,\authormark{1} Hong-Mei Zou,\authormark{1,*} Jianhe Yang,\authormark{1} and Danping Lin\authormark{2}}

\address{\authormark{1}Department of Physics, Key Laboratory for Low-Dimensional Structures and
	Quantum Manipulation (Ministry of Education), and Synergetic Innovation
	Center for Quantum Effects and Applications of Hunan, Hunan Normal
	University, Changsha 410081, People's Republic of China\\
\authormark{2}Faculty of Science, Guilin University of Aerospace Technology, Guilin 541004,
People's Republic of China}

\email{\authormark{*}zhmzc1997@hunnu.edu.cn}

\begin{abstract}
In this paper, a two-level atom coupled with double Lorentzian spectrum is solved by pseudomode theory, and an analytic representation of the density operator is obtained. 
Secondly, the paper investigate the entanglement witness and entropy uncertainty, and get the analytical representation of entanglement and uncertainty and their relationship. The environmental effects of the double Lorentzian spectrum are explained by non-Markovianity. In addition, this paper study the influence of Zeno effect on entanglement witness and uncertainty. The results show that the Zeno effect not only can effectively prolong the time of entanglement witness and reduce the lower bound of the entropy uncertainty, but also can greatly enhance the time of entanglement witness and reduce the entanglement value of witness.
\end{abstract}

\section{INTRODUCTION}
Heisenberg uncertainty principle constitutes the basic element of quantum mechanics\cite{Berta,Dupuis,Koenig.2012,Jarzyna,DiVincenzo,Cerf,Grosshans,Vallone,Chitambar.2019,Ma.2016,Watrous.2018,Heisenberg}. Then Kennard and Robertson
used standard variance to represent the uncertainty relationship between two incompatible observations\cite{Kennard,Robertson}, i.e., $\Delta {Q}\cdot \Delta {R}\geq \frac{1}{2}|\langle \lbrack \hat{Q},\hat{R}]\rangle |$, 
but it will result in a mediocre consequence if the $[\hat{Q},\hat{R}]=0$ or $\langle \lbrack \hat{Q},\hat{R}]\rangle =0$.
In order to overcome this shortcoming, a new expression
called entropy uncertainty relation(EUR) is proposed by Deutsch
and improved by Kraus, and then is demonstrated by
Maassen and Uffink\cite{Deutsch,Kraus,Maassen1988}. Until recently, Berta \textit{et al.} proposed a quantum-memory-assisted entropy uncertainty relationship\cite{Berta}, which is given by

\begin{equation}\label{E2}
\begin{split}
S(P|B)+S(Q|B)\geq \log_{2}\frac{1}{c}+S(A|B).
\end{split}
\end{equation}

Where $S(A|B)=S\left( \hat{\rho}_{AB}\right) -S\left( \hat{\rho}_{B}\right) $
is the conditional von Neumann entropy, $\hat{\rho}_{B}$ 
is the reduced density matrix of the 
$\hat{\rho}_{AB}$. In the same way, $S(P|B)\left(S(Q|B)\right) $ is the conditional entropy of the post-measurement state $\hat{\rho}_{PB}\left(\hat{\rho}_{QB}\right) $ 
for which atom A measures $P\left( Q\right) $. The entropy uncertainty relation was originally put forward to solve the conceptual shortcomings in the uncertainty principle, so it plays an important role in the quantum foundation. Prevedel \textit{et al.} and Li \textit{et al.} experimentally demonstrated the quantum-memory-assisted entropy uncertainty relation\cite{Prevedel,Li}. 
Justin and Franco also suggested alternative definitions of error and disturbance then these definitions naturally produce complementarity and error-disturbance inequalities that had the same form as the traditional Heisenberg relation\cite{Justin2014}. It can raise the likelihoods of future measurements on quantum systems.
Likewise, it is providing the foundation for the security of many quantum cryptographic protocols. For instance, security proofs in the noisy-storage model are intimately connected to entropy uncertainty relation\cite{Koenig.2012,Wehner.2008}. In addition, entanglement witnessing is a well-developed field\cite{Guhne.2009,F.B.M.DosSantos.2013}. Here, we focus mostly on entanglement witnesses that follow from entropy uncertainty relation.

On the other hand, quantum entanglement is an important quantum
resource in quantum information processing tasks such as quantum teleportation, quantum dense coding, quantum cryptography and quantum computing\cite{Amico,Bouwmeester.1997,Bouwmeester.2000,Jennewein,Jozsa}. For the past years, the research on generating steady state entanglement has attracted extensive attention\cite{Adam2015,Yueh2013}. For example, Ye-hong Chen \textit{et al.} explored an interesting alternative for a fast and high-fidelity generation of steady state entanglement\cite{Ye2019}. Based on the above research, Wei and Adam \textit{et al.} also demonstrated the generation of steady-state nearly maximal quantum entanglement\cite{Wei2018}.
However, quantum
entanglement is easily destroyed because any interaction between quantum systems and their surroundings can give rise to decoherence and dissipation phenomena. Hence, how to effectively protect and witness quantum entanglement of open quantum systems have attracted wide attention during the last decade\cite{Breuer,Duan.2000,Weiss,Yu}. For example, Ming-Liang Hu and Heng Fan   investigated entanglement protection and entanglement witness of open systems using the entropy uncertainty relation in the presence of quantum memory\cite{Hu}. The authors in Ref.\cite{Zou} studied tripartite disentangling and entangling dynamics as well as protecting bipartite entanglement with both
atom-atom interactions and atom-cavity couplings taken simultaneously into account. 

As a matter of fact, many methods have been developed to extend the entanglement witness time of a quantum system. Particularly quantum Zeno effect(QZE) is often used to protect entanglement and prolong the time of entanglement witness via frequent measurements\cite{Francica.2010,Maniscalco.2008,Petrosky1991,Qing2013,Lan2009}. For instance, the authors studied the possibility of modifying the dynamics of both quantum correlations, such as entanglement, discord, and classical correlations of an open bipartite system by means of the quantum Zeno effect\cite{Francica.2010}.

In this paper, we focus on a two-atom system under the quantum Zeno effect, in which each atom is in an environment with a double Lorentzian spectrum, and there is no interaction between the two subsystems. Pseudomode theory is used to solve this model and an equation in Lindblad form is obtained, from which we find the dissipation and coupling terms related to the double Lorentzian spectrum. Under pseudomode theory, the interaction between each atom and its environment is replaced by the interaction between the atom and pseudomodes, that is, each subenvironment is equivalent to two pseudomodes. The pseudomodes act as a memory for the atom. Then we investigate entanglement witness and entropy uncertainty of the open two-atom system. Besides, we analyze influence of quantum Zeno effect and environmental parameters on entanglement witness and entropy uncertainty. Finally, to explain the total effect of the pseudomodes on each atom, we introduce the non-Markovianity. The results show that appropriate parameters and quantum Zeno effect can effectively reduce the lower bound of the entropy uncertainty and prolong the time of entanglement witness. 

Our work has the following features: First, the model we consider is widely studied in
the theoretical studies of dynamics of open quantum systems,
and the double Lorentzian spectrum is used to describe the influence of the environment to the system more realistically. And the non-Markovianity are used to explain the physical meaning of the double Lorentzian spectrum. Second, We get a concrete, brief entanglement expression, the lower bound of the entropy uncertainty expression, and relationship of them in this model.

The paper is organized as follows. In Sec.II, we give a physical model and its solution. In Sec.III, we review quantum entanglement, entropy uncertainty relation and Zeno effect with quantum memory. In Sec.IV, we calculate entanglement witness and entropy uncertainty with and without Zeno effect, respectively. We also discuss their physical explanation. In Sec.V, we summarize our conclusions.

\section{PHYSICAL MODEL}
Firstly, we consider a
two-level atom coupling with a zero-temperature environment. Assuming $\hbar =1$, the total Hamiltonian of the system is given by
\begin{equation}\label{E3}
\begin{split}
\hat{H}=\hat{H}_{0}+\hat{H}_{I},
\end{split}
\end{equation}
where
\begin{equation}\label{E4}
\begin{split}
\hat{H}_{0}=\omega _{0}\hat{\sigma }_{+}\hat{\sigma }_{-}+\sum_{k}\omega _{k}%
\hat{b}_{k}^{\dagger}\hat{b}_{k},
\end{split}
\end{equation}
and
\begin{equation}\label{E5}
\begin{split}
\hat{H}_{I}=\sum_{k}(g_{k}\hat{b}_{k}\hat{\sigma }_{+}+g_{k}^{\ast }\hat{b}_{k}^{\dagger}\hat{\sigma }_{-}),
\end{split}
\end{equation}
$\hat{\sigma}_{+}$ and $\hat{\sigma}_{-}$ are the raising and lowering operators for the atom with the transition frequency $\omega _{0}$. $\hat{b}_{k}^{\dagger }$ and $\hat{b}_{k}$ are the creation and annihilation
operators for the $k$-th  mode of the environment with frequency $\omega _{k}$. Also, the strength of the coupling between the atom and the $k$-th mode of the environment is given by $g_{k}$. For convenience in the later discussion, we consider the coupling coefficient is real. 
\begin{figure}\label{EP}
	\centering\includegraphics[width=0.6\textwidth]{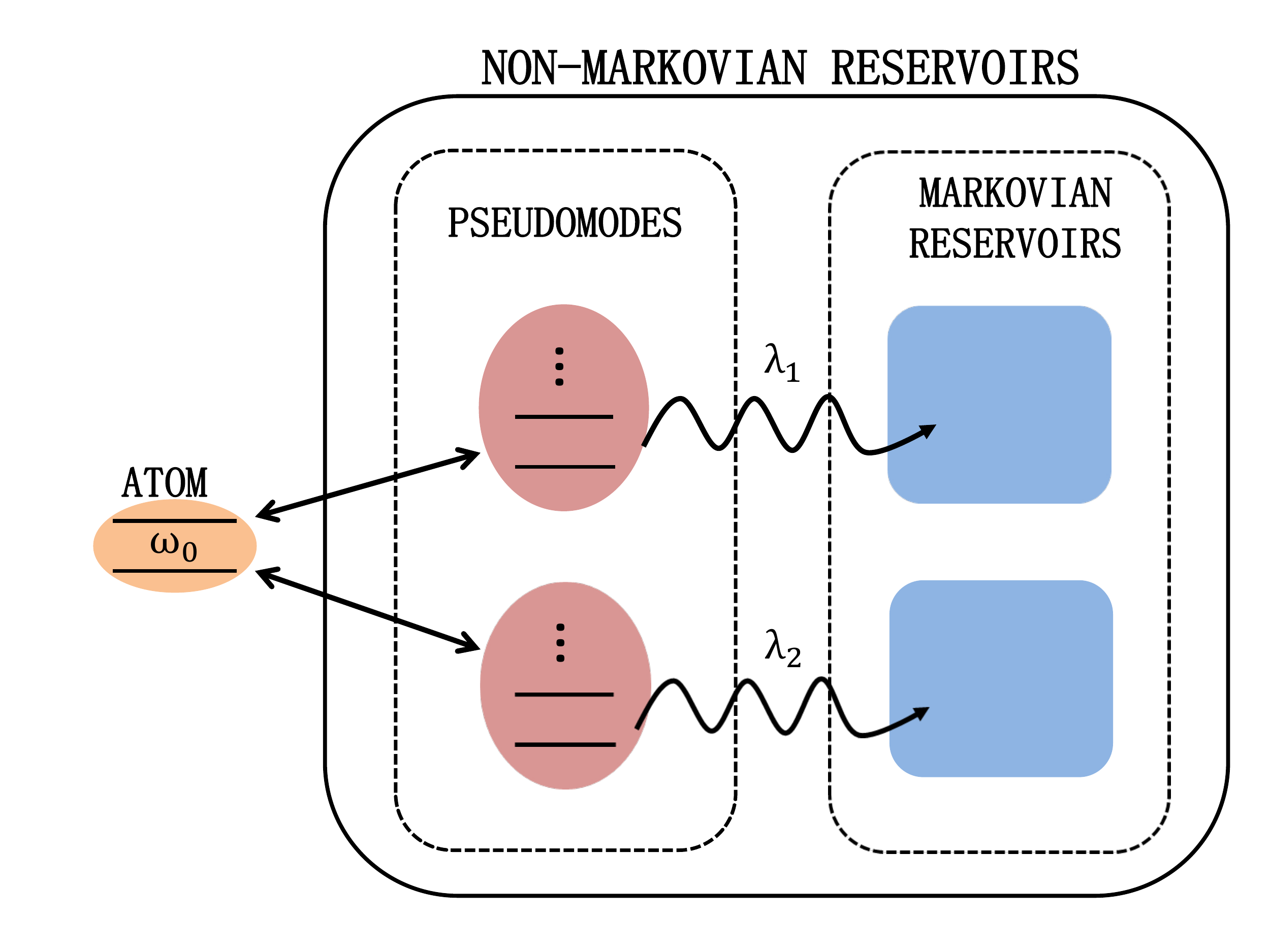}
	\caption{Schematic illustration of the atom interactions with Lorentzian environment which is simply a sum of two Lorentzian functions. The interaction between atom and the pseudomodes replaces the interaction it and environment.}
\end{figure}

For a single excitation of the total system,
we suppose that the initial state is
\begin{equation}\label{E6}
|\Phi \left( 0\right) \rangle =M_{0}(0)|g\rangle |0\rangle
_{R}+M_{1}\left( 0\right) |e\rangle |0\rangle _{R}.
\end{equation}
We will now expand a general state vector of the total system as
\begin{equation} \label{E7}
\begin{aligned} |\Phi \left( t\right) \rangle =&M_{0}(t)|g\rangle
|0\rangle _{R}+M_{1}\left( t\right) |e\rangle|0\rangle _{R}\\
&+\sum_{k}M_{k}\left( t\right) |g\rangle |1_{k}\rangle _{R}.
\end{aligned}
\end{equation}
By substituting Eq.\eqref{E7} into the \bigskip Schr\"{o}dinger
equation, the integrodifferential equation satisfied by the probability amplitude can be obtained
\begin{equation}\label{E8}
\frac{dM_{1}\left( t\right) }{dt}=-\int_{0}^{t}f\left( t-t^{\prime }\right)
M_{1}\left( t^{\prime }\right) dt^{\prime }.
\end{equation}
Where the correlation function is related to the spectral density $J\left(\omega \right) $ of the environment by
\begin{equation}\label{E9}
f\left( t-t^{\prime }\right) =\int_{0}^{\infty }J\left( \omega \right)e^{i\left( \omega _{0}-\omega \right) \left( t-t^{\prime }\right) }d\omega.
\end{equation}

For different environments, there are several different forms of spectral density, e.g., the Lorentzian spectrum and the Ohmic spectrum are usually used to simulate the environment in cavity QED system\cite{Behzadi.2017,Maniscalco.2008,Fanchini.2013}. However, for band-gap environments, the double Lorentzian spectrum can describe the influence of the environment on the system more realistically\cite{Calajo.2017}. The double Lorentzian spectral density is considered in this paper, i.e.,

\begin{equation}\label{E10}
J\left( \omega \right) =\frac{1}{2\pi }\left[ \frac{\gamma _{1}\lambda_{1}^{2}}{\left( \omega -\omega _{0}\right)^{2}+\lambda _{1}^{2}}+\frac{\gamma _{2}\lambda _{2}^{2}}{\left( \omega -\omega _{0}\right) ^{2}+\lambda
	_{2}^{2}}\right].
\end{equation}%
Here, in order to solve $f(t-t')$ in Eq.\eqref{E9} from Eq.\eqref{E10}, we will use the pseudomode theory\cite{Garraway.1997,Garraway.1996,Mazzola.2009}. The psudomodes is determined by the singularity of the spectral density in the lower half plane and possesses the properties of a finite-Q cavity mode. The interaction between the atom and the environment can be replaced by the interaction between the atom and the psudomodes. Diagrammatic representation of the atom-pseudomode dynamics in Fig.1.

The constant$\ \omega _{0}$ is the oscillation frequency of the
psudomodes, the parameter$\ \lambda _{j}\left( j=1,2\right) $ is the decay rate of the $j$-th pseudomode. The constants
$\omega _{0}$ and $\lambda_{j}$ are depend on the position of the pole $z_{j}=$ $\omega _{0}-i\lambda_{j}$ while $\bar{g}_{j}=\sqrt{\frac{\gamma _{j}\lambda _{j}}{2}}$ is the 
coupling strength between the atom and $j$-th pseudomode. In the weak coupling regime, that is, for 
$\lambda _{j}>2\gamma _{j}(j=1,2)$ the behavior of dynamical evolution of the system coupling to the j-th pseudomode is essentially a Markovian exponential decay controlled by 
$\lambda _{j}\left( j=1,2\right)$. 
Instead, in the strong coupling regime, that is , for $\lambda _{1}<2\gamma
_{1}(j=1,2)$, the evolution is non-Markovian(See appendix A for equation in Lindblad form).

Substituting $J(\omega)$ into Eq.\eqref{E9}, we get
\begin{equation}\label{E11}
f\left(t-t^{\prime}\right)=\frac{1}{2}\left(\gamma_{1}\lambda_{1}e^{-iz_1(t-t^{\prime })}+\gamma_{2}\lambda_{2}e^{-iz_2(t-t^{\prime
	})}\right),
\end{equation}
with this result inserted into Eq.\eqref{E8} and move to an interaction representation, we will find that
\begin{equation}\label{E12}
\begin{aligned}
i\dot{M}_{1}(t)=&-i\int_{0}^{t}(\bar{g}_{1}^2 e^{-iz_{1}(t-t^{\prime})}\\
&+\bar{g}_{2}^2e^{-iz_{2}
	(t-t^{\prime})})M_{1}\left( t^{\prime }\right) dt^{\prime }\\
=&\bar{g}_{1}\bar{P}_1(t)+\bar{g}_{2}\bar{P}_2(t),\\
i\dot{\bar{P}}_{1}(t)=&z_{1}\bar{P}_{1}(t)+\bar{g}_{1}M_{1}(t),\\
i\dot{\bar{P}}_{2}(t)=&z_{2}\bar{P}_{2}(t)+\bar{g}_{2}M_{1}(t),
\end{aligned}
\end{equation}
where $\bar{P}_j(t)=-i\bar{g}_{j}\int_{0}^{t}e^{-iz_{j}
	(t-t^{\prime})}M_{1}\left( t^{\prime }\right) dt^{\prime }$ is the $j$-th pseudomode amplitude.

By using the Laplace transform of $M_1(t)$, we can find that 
\begin{equation}\label{E13}
M_{1}(s) =\frac{A}{B}M_{1}(0),
\end{equation}
where $A=\left( s+\lambda _{1}\right)\left( s+\lambda _{1}\right)$
and $B=s^{3}+s^{2}\left( \lambda _{1}+\lambda _{2}\right) +s\left( \lambda
_{1}\lambda _{2}+\frac{1}{2}\gamma _{1}\lambda _{1}+\frac{1}{2}\gamma
_{2}\lambda _{2}\right) +\frac{1}{2}\lambda _{1}\lambda _{2}\left( \gamma
_{1}+\gamma _{2}\right)$.

Thus, we can get the analytical solution of the probability amplitude $M_{1}\left( t\right)$ by inverse Laplace transformation, namely,
\begin{equation}\label{E14}
M_{1}\left(t\right)=\sum_{i=1}^{3}Res_{s=s_i}[M_{1}(s)e^{st},s_i],
\end{equation}
where $s_{i}$ are the three solutions of B=0 of Eq.\eqref{E13}.

Consequently, the density matrix is written as 
\begin{equation}\label{E15}
\hat{\rho}_{S}\left( t\right) =\left(
\begin{array}{cc}
|M_{1}\left( t\right) |^{2} & M_{1}\left( t\right) M_{0}^{\ast } \\
M_{0}M_{1}^{\ast }\left( t\right)  & 1-|M_{1}\left( t\right) |^{2}
\end{array}\right) . 
\end{equation}

Next, we consider a two-atom system formed by two identical subsystems that do not interact with each other. Each subsystem consists of an atom, which interacts with the environment locally. The density matrix $\hat{\rho }^{\Phi }(t)$ of two-atom system can be determined by the procedure presented in Ref.\cite{ Bellomo}. 

In the standard basis $\mathcal{A}=\left\{ |1\rangle \equiv |ee\rangle \text{%
	, }|2\rangle \equiv |eg\rangle \text{, }|3\rangle \equiv |ge\rangle \text{, }%
|4\rangle \equiv |gg\rangle \right\} $. The environment R is initially at $|0\rangle_{R}$. The initial state of the atom is set as 
$|\Phi \left( 0\right) \rangle =
\frac{1}{\sqrt{2}}\left[ |eg\rangle +|ge\rangle \right] $, the $|e\rangle$ and $|g\rangle$ is the excited state and ground state, respectively. We obtain the diagonal and off-diagonal elements of the reduced density matrix at the initial time,
\begin{equation}  \label{E16}
\hat{\rho }_{22}^{\Phi }\left( 0\right) =\hat{\rho }_{23}^{\Phi }\left(
0\right) =\hat{\rho }_{32}^{\Phi }\left( 0\right) =\hat{\rho }_{33}^{\Phi
}\left( 0\right) =\frac{1}{2},
\end{equation}
and everything else is equal to $0$.

After time $t>0$, the elements of the reduced density matrix can be written as
\begin{subequations}
	\begin{align}
	\hat{\rho }_{22}^{\Phi }\left( t\right) &=|M_{1}\left( t\right) |^{2} \hat{\rho }_{22}^{\Phi}\left( 0\right),\\
	\hat{\rho }_{33}^{\Phi }\left( t\right) &=|M_{1}\left( t\right) |^{2} \hat{\rho }%
	_{33}^{\Phi }\left( 0\right),\\
	\hat{\rho }_{23}^{\Phi }(t) &=|M_{1}\left( t\right) |^{2}\hat{\rho}_{23}^{\Phi }( 0),\\
	\hat{\rho }_{44}^{\Phi }\left( t\right) &=( 1-|M_{1}\left( t\right) |^{2})(\hat{\rho }_{22}^{\Phi }(0) +\hat{\rho }_{33}^{\Phi }(0)).
	\end{align}
\end{subequations}
It's obvious that this is an X structure density matrix\cite{Bose.2001,Pratt2004,Hagley.1997}:

\begin{equation}  \label{Edensity}
\hat{\rho }^{\Phi }\left( t\right) =\left(
\begin{array}{cccc}
0 & 0 & 0 & 0 \\
0 & \hat{\rho }_{22}^{\Phi }\left( t\right) & \hat{\rho }_{23}^{\Phi }\left(
t\right) & 0 \\
0 & \hat{\rho }_{32}^{\Phi }\left( t\right) & \hat{\rho }_{33}^{\Phi }\left(
t\right) & 0 \\
0 & 0 & 0 & \hat{\rho }_{44}^{\Phi }\left( t\right)%
\end{array}
\right).
\end{equation}

\section{QUANTUM ENTANGLEMENT, EUR and ZENO EFFECT WITH QUANTUM MEMORY}
\subsection{Quantum Entanglement}

There are many methods to measure quantum entanglement, for
examples, the relation of entropy entanglement, the partial entropy
entanglement and the concurrency. The entanglement dynamics of the two-atom system can be get by using Wootters concurrence\cite{Wootters.1998}. The concurrence is derived from the reduced density matrix of the two-atom
systems as%
\begin{equation}  \label{concurrence}
C_{\hat{\rho }}\left( t\right) =\max \left\{ 0,\sqrt{\beta _{1}}-\sqrt{%
	\beta _{2}}-\sqrt{\beta _{3}}-\sqrt{\beta_{4}}\right\},
\end{equation}
where $\beta _{i}$ is the eigenvalue of matrix $\tilde{\rho }$ in
decreasing order.
\begin{equation}  \label{E20}
\tilde{\rho }=\hat{\rho }^{\Phi }\left( t\right) \left(\hat{\sigma}
_{y}\otimes \hat{\sigma} _{y}\right) \hat{\rho }^{\Phi }\left( t\right)
^{\ast }\left( \hat{\sigma} _{y}\otimes \hat{\sigma} _{y}\right).
\end{equation}

It is well-known that $\hat{\sigma} _{y}$ is Pauli matrices and $\hat{\rho }%
^{\Phi }\left( t\right) ^{\ast }$ is the complex conjugation of $\hat{\rho }%
^{\Phi }\left( t\right) $ in the standard basis $\mathcal{A}$. Concurrency $%
C_{\hat{\rho }}\left( t\right) $ ranges from 0 to 1, representing
disentangled state to maximally entangled state. Under our dynamical conditions, the X density matrix is keep during the two-atom system evolution. Hence, we can obtain the concurrency from Eq.\eqref{Edensity} to Eq.\eqref{concurrence},
\begin{equation}  \label{Eq21}
C_{\hat{\rho }}(t) =2\max\{0,K(t)\},
\end{equation}
where $K(t)=|\rho_{23}(t)|-\sqrt{\rho_{11}(t)\rho_{44}(t)}$. Using the above expressing, we calculate that the concurrency is 
\begin{equation}  \label{Eq2}
C_{\hat{\rho }}(t) =|M_{1}\left( t\right) |^{2}=\langle \Phi(0) |\hat{\rho }^{\Phi }(t)|\Phi
(0) \rangle.
\end{equation}
The Eq.\eqref{Eq2} shows that the concurrency is directly dependent on the survival probability of the initial state.

\subsection{Entropy Uncertainty Relation(EUR)}

According to Eq.\eqref{E2}, and we order $\hat{P}=\hat{S}_x,\hat{Q}=\hat{S}_y$
\begin{equation}\label{EUR}
S(\hat{S}_{x}|B)+S(\hat{S}_{y}|B)\geq \log_{2}\frac{1}{c}+S(A|B). 
\end{equation}

Once measures $\hat{S}_{x}$ or $\hat{S}_{y}$, the two-atom state is
\begin{equation}\label{E23}
\hat{\rho}_{S_{x}B}=\sum_{i}\left( |\psi _{l}\rangle \langle \psi
_{l}|\otimes I_{B}\right) \hat{\rho}_{AB}\left( |\psi _{k}\rangle \langle
\psi _{k}|\otimes I_{B}\right) , 
\end{equation}

or
\begin{equation}  \label{E24}
\hat{\rho }_{S_{y}B}=\sum_{i}\left( |\phi _{l}\rangle \langle \phi
_{l}|\otimes I_{B}\right) \hat{\rho} _{AB}\left( |\phi _{k}\rangle \langle
\phi _{k}|\otimes I_{B}\right).
\end{equation}%
$S\left( \hat{\rho }_{j}\right) =-tr\left( \hat{\rho }_{j}\log_{2}\hat{\rho }%
_{j}\right) =-\sum_{i}\alpha _{i}\log_{2}\alpha _{i}$ is the von Neumann entropy, $\alpha
_{i}$ is the eigenvalues of the density matrix $\hat{\rho }_{j}$. $c\equiv
\max_{l,k}|\langle \psi _{l}|\phi _{k}\rangle |^{2}=\frac{1}{2}$ is defined as the maximum complementary for two incompatible observable $\hat{S}_{x}$ and $\hat{S}_{y}$. $|\psi _{l}\rangle $ and $|\phi _{k}\rangle $ are the eigenstates of
the $\hat{S}_{x}$ and $\hat{S}_{y}$, respectively.

The left-hand side of inequality Eq.\eqref{EUR} represents the entropy uncertainty(EUR) about atom B after measuring
atom A, i.e., $EUR=S(\hat{S}_{x}|B)+S(\hat{S}_{y}|B)$. The right-hand side of the inequality equation \eqref{EUR} provides
the lower bound of the entropy uncertainty(LEU), i.e., $LEU=\log_{2}\frac{1}{c}+S(A|B)$. 
We can calculate that $LEU$ is equal to
\begin{equation}  \label{E26}
\begin{aligned}
LEU=&-(1-|M_{1}|^2)\log_{2}(1-|M_{1}|^2)-|M_{1}|^2\log_{2}|M_{1}|^2\\
&+(1-\frac{1}{2}|M_{1}|^2)\log_{2}(1-\frac{1}{2}|M_{1}|^2)\\
&+\frac{1}{2}|M_{1}|^2\log_{2}\frac{1}{2}|M_{1}|^2+1\\
=&-(1-C_{\hat{\rho }}(t))\log_{2}(1-C_{\hat{\rho }}(t))-C_{\hat{\rho }}(t)\log_{2}C_{\hat{\rho }}(t)\\
&+(1-\frac{1}{2}C_{\hat{\rho }}(t))\log_{2}(1-\frac{1}{2}C_{\hat{\rho }}(t))\\
&+\frac{1}{2}C_{\hat{\rho }}(t)\log_{2}\frac{1}{2}C_{\hat{\rho }}(t)+1.
\end{aligned}
\end{equation}
It's worth noting that from equation \eqref{E26} you can intuitively see that when $C_{\hat{\rho }}(t)$ is equal to $0$, $LEU$ is equal to $1$, and $C_{\hat{\rho }}(t)$ is equal to $1$, $LEU$ is equal to $0$.

\subsection{Zeno Effect}

The quantum Zeno effect is the inhibition of transitions between quantum states by frequent measurements. Namely, an unstable quantum system will not decay if it is measured frequently. It has been discussed theoretically\cite{Koshino.2005,Facchi.2001,Maniscalco.2006} as well as experimentally\cite{Bernu.2008,Lan2009}. In the
other word, the survival probability $P_{N}\left( t\right) $ of the system
in its state $|\Phi _{S}\rangle $ remains constant in the limit $N\rightarrow\infty $, which suppose $N$ measurements with equal time interval $T=\frac{t}{N}$.
The initial state survival probability is given by\cite{Maniscalco.2008}
\begin{equation}\label{E28}
\begin{aligned} P_{N}\left(t\right)&=\left[P\left(T\right)\right]^{N}\\
&=\langle \Phi\left( 0\right) |\hat{\rho }^{\Phi }\left( T\right) |\Phi
\left(0\right) \rangle^{N}, 
\end{aligned} 
\end{equation}%
where time $t=NT$, $P\left( T\right) $ is the probability of system in the initial state
right after a measurement is performed$\left( N=1\right) $. Then assumption
that the environment is reset to its initial state $\hat{\rho}_{E}\left(
0\right) $ is implicitly made so that the evolution of two-atom system
in the next Zeno interval is the same as that in the previous interval, and
consequently $P_{N}\left( t\right) =\left[ P\left( T\right) \right] ^{N}$.

\section{RESULT AND DISCUSSION}
\subsection{ Entanglement Witness and Entropy Uncertainty without Zeno Effect}
\begin{figure*}
	\centering\includegraphics[width=0.49\textwidth]{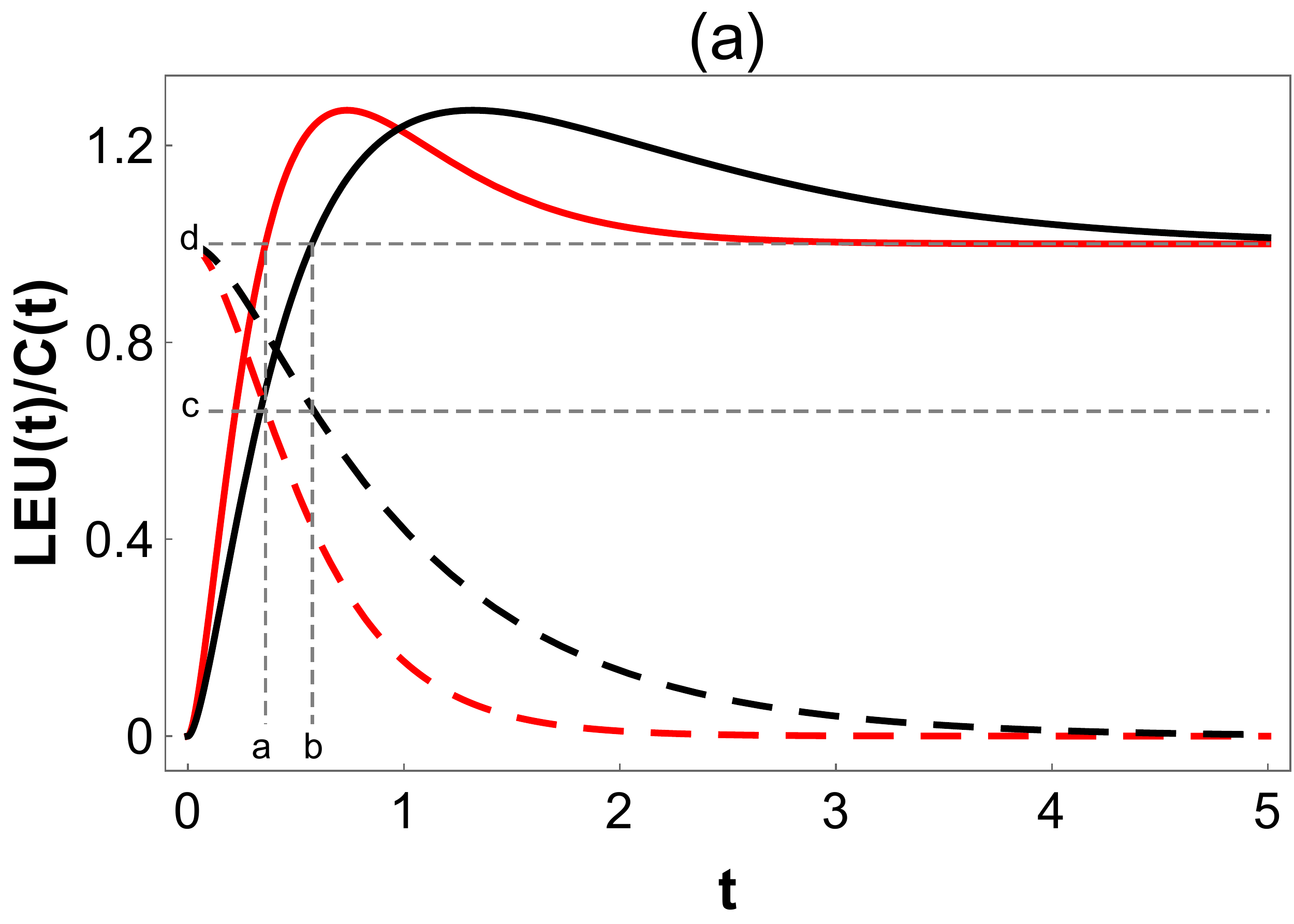}
	\centering\includegraphics[width=0.49\textwidth]{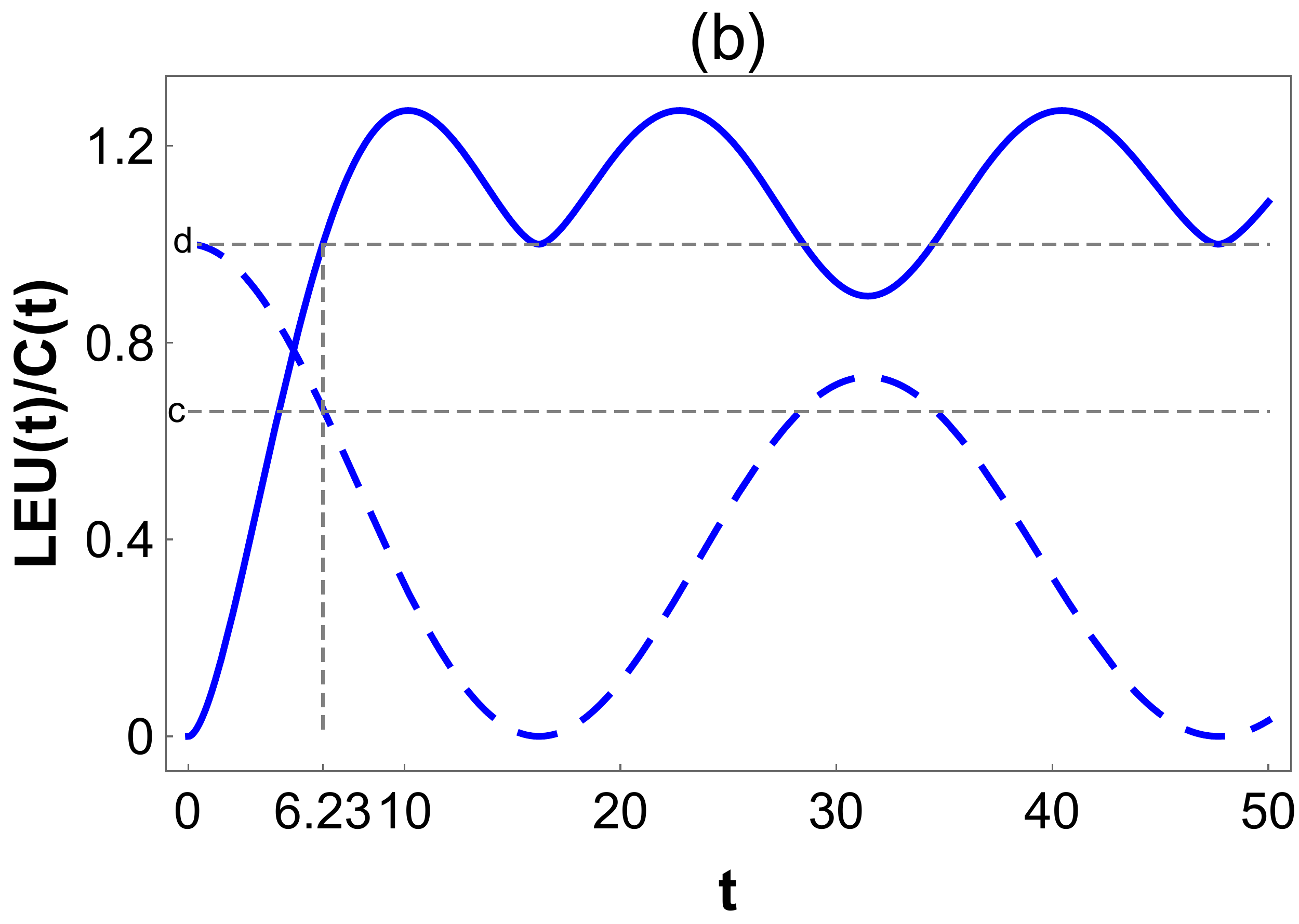}
	\caption{Time evolution of the $LEU(t)$(solid) and $C(t)$(dashed) for the different type.$(a)$ $\lambda _{1}=5\gamma _{1}$, $\lambda _{2}=5\gamma_{2}$(red line), $\lambda _{1}=5\gamma _{1}$, $\lambda _{2}=0.01\gamma_{2}$(black line); $(b)$ $\lambda _{1}=0.01\gamma _{1}$, $\lambda _{2}=0.01\gamma_{2}$(blue line). All types are at $\gamma_{1}=\gamma_{2}$. Here $a=0.36, b=0.576, c=0.66, d=1$. }
\end{figure*}

We now look at the entanglement dynamics($C(t)$) and the lower bound of the entropy
uncertainty relationship($LEU(t)$) for the weak coupling and strong coupling,
i.e., for $\lambda >2\gamma $ and $\lambda <2\gamma $, respectively. In Fig. 2(a) for $\lambda _{1}=5\gamma _{1}$, $\lambda _{2}=5\gamma_{2}$, the atom is weak coupling with two pesudomodes, namely, the information of the atom quickly dissipates to the two pesudomodes. Therefore, we can see that the $C(t)$ decreases monotonically to zero, and the $LEU(t)$ monotonically increase and then quickly tend to 1. The entanglement witness time $\tau$ is $0.36$. For $\lambda _{1}=5\gamma _{1}$, $\lambda _{2}=0.01\gamma _{2}$, the atom is weak coupling with $1$-th pesudomode and strong coupling with $2$-th pesdomode. Namely, the information of the atom only quickly dissipates to the $1$-th pesudomode, and the information of the $2$-th pesudomode will flow back to the atom. Therefore, the image of $C(t)$ is similar to the former. The peak of uncertainty
relationship is the same. Only the rate of change of the latter is slower
than that of the former. Similarly, the image of $LEU(t)$ is resemble the
former. Besides, the rate of decline is smaller than the former. And the entanglement witness time $\tau$ is $0.576$. That is because the information going back from the $2$-th pseudomode to the atom, so $\tau$ is greater than the previous case. In both cases, there is no entanglement recovery. Whereas, in Fig. $2(b)$ for $\lambda _{1}=0.01\gamma _{1}$, 
$\lambda_{2}=0.01\gamma _{2}$, the atom is strong coupling with two pesudomodes. Not only does the decay of atomic information slow down, but also the information will flow back to the atom from the two pesudomodes due to the memory and feedback of the pesudomodes.  For $LEU(t)$, the rate of monotonically increase is smaller. The $\tau$ is $6.23$. In the other word, the stronger the coupling between the atom and the pseudomodes, the longer the entanglement witness time, and the greater the $C(t)$. It is noteworthy that for the case where the atom is strongly coupled to both pseudomodes, a second entanglement witness period occurs.This indicates that more information is coming back to the system in strong coupling. But there is the same minimal value$(c=0.66)$ of entanglement witness in the three cases, i.e. $\lambda _{1}=5\gamma _{1}$, $\lambda _{2}=5\gamma_{2}$, and $\lambda _{1}=5\gamma _{1}$, $\lambda _{2}=0.01\gamma_{2}$ and $\lambda _{1}=5\gamma _{1}$, $\lambda _{2}=5\gamma_{2}$.

\subsection{The Degree of Non-Markovianity}
\begin{figure}\label{EqN}
	\centering\includegraphics[width=0.6\textwidth]{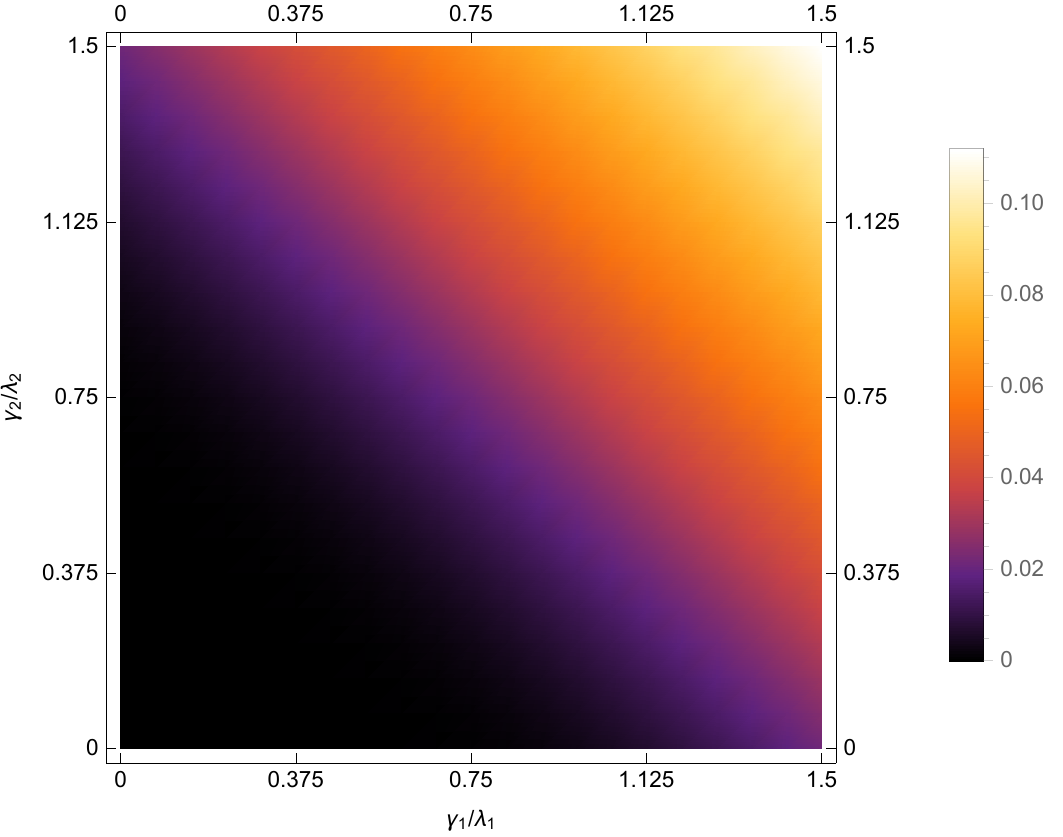}
	\caption{(Color online)Contour map of the  
		non-Markovianity of the two-atom system.$0<\gamma_{i}<0.5(i=1,2)$ is Markovian, and $0.5<\gamma_{i}(i=1,2)$ is non-Markovian.}
\end{figure}

In order to explain the above results in more detail, we introduce the non-Markovianity \cite{Kang2019,Wei2012,Breuer2010}. Due to the initial state is a X density in this paper, we can straightforward obtain the non-Markovianity of the two-atom system\cite{Fanchini.2013}. On the other hand, Zhi He $et\ al.$ found that, for phase damping and amplitude damping channels, the three previous measures of non-Markovianity is equivalent, i.e., the measures based on the
dynamical divisibility, quantum trace distance, and quantum mutual information\cite{He.2017}.
Now we measure the non-Markovianity by using the method based on quantum
trace distance. We get the non-Markovianity for the single atom system firstly, and then we generalize it to the two-atom system. The non-Markovianity is mathematically defined as follows:
\begin{equation}  \label{Eq28}
\mathcal{N}_{1}=\max\limits_{\hat{\rho}_{1}(0),\hat{\rho}_{2}(0)}\int_{[dD_{12}(t)/dt]>0}\frac{dD_{12}(t)}{dt}dt,
\end{equation}
where the trace distance $D_{12}(t)=\frac{1}{2}\text{tr}|\hat{\rho}_{1}(t)-\hat{\rho}_{2}(t)|$ between two reduced density matrices $\hat{\rho}_{1}(t)$ and $\hat{\rho}_{2}(t)$. We taking $\hat{\rho}_{1}(0)=|g\rangle\langle{g}|$ and $\hat{\rho}_{2}(0)=|e\rangle\langle{e}|$. According to the literature\cite{Fanchini.2013}, we can get that the non-Markovianity of the two-atom system is $\mathcal{N}=2\mathcal{N}_{1}$. We plot the non-Markovianity of the two-atom system.

\begin{figure*}
	\centering\includegraphics[width=0.31\textwidth]{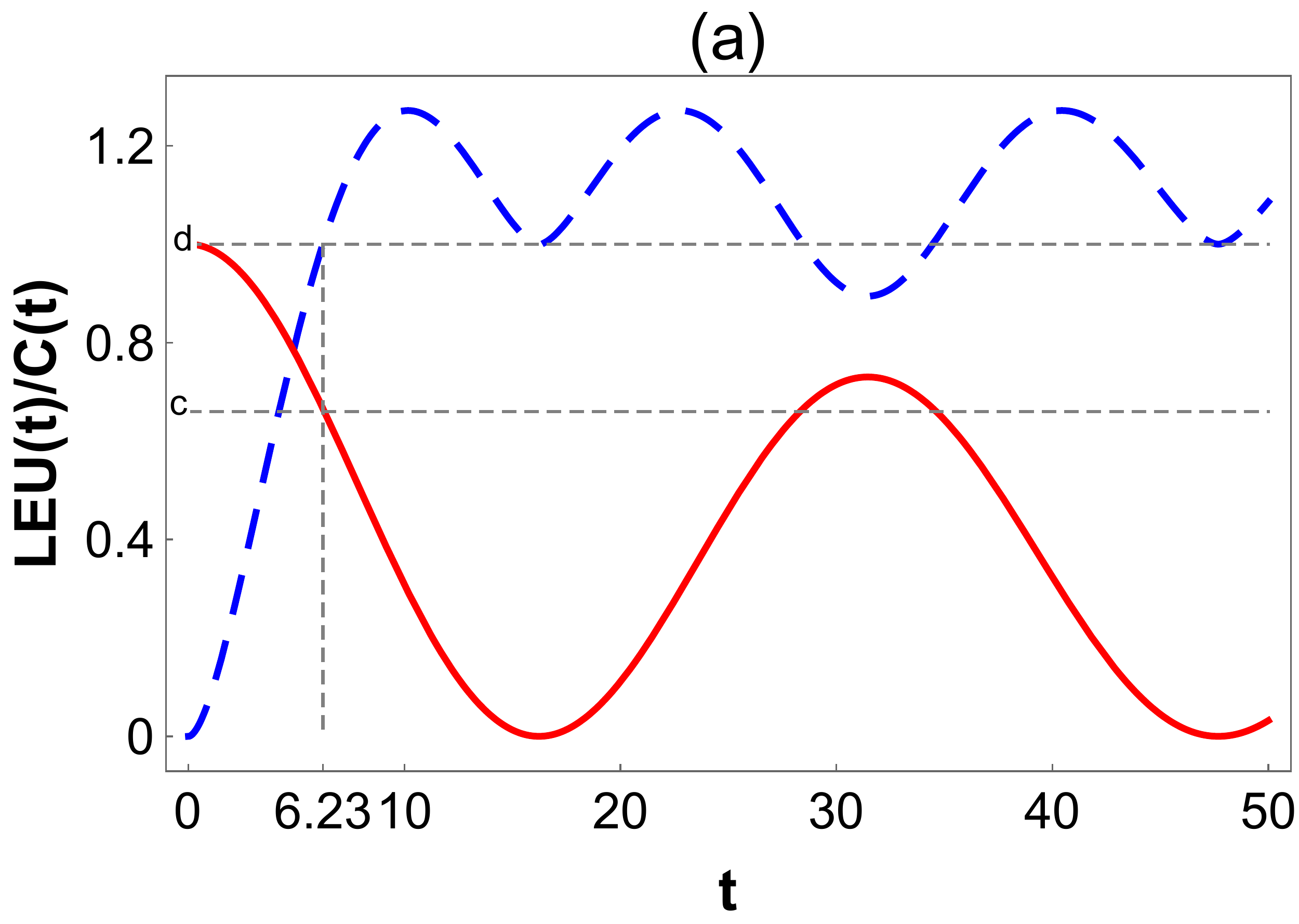}
	\centering\includegraphics[width=0.31\textwidth]{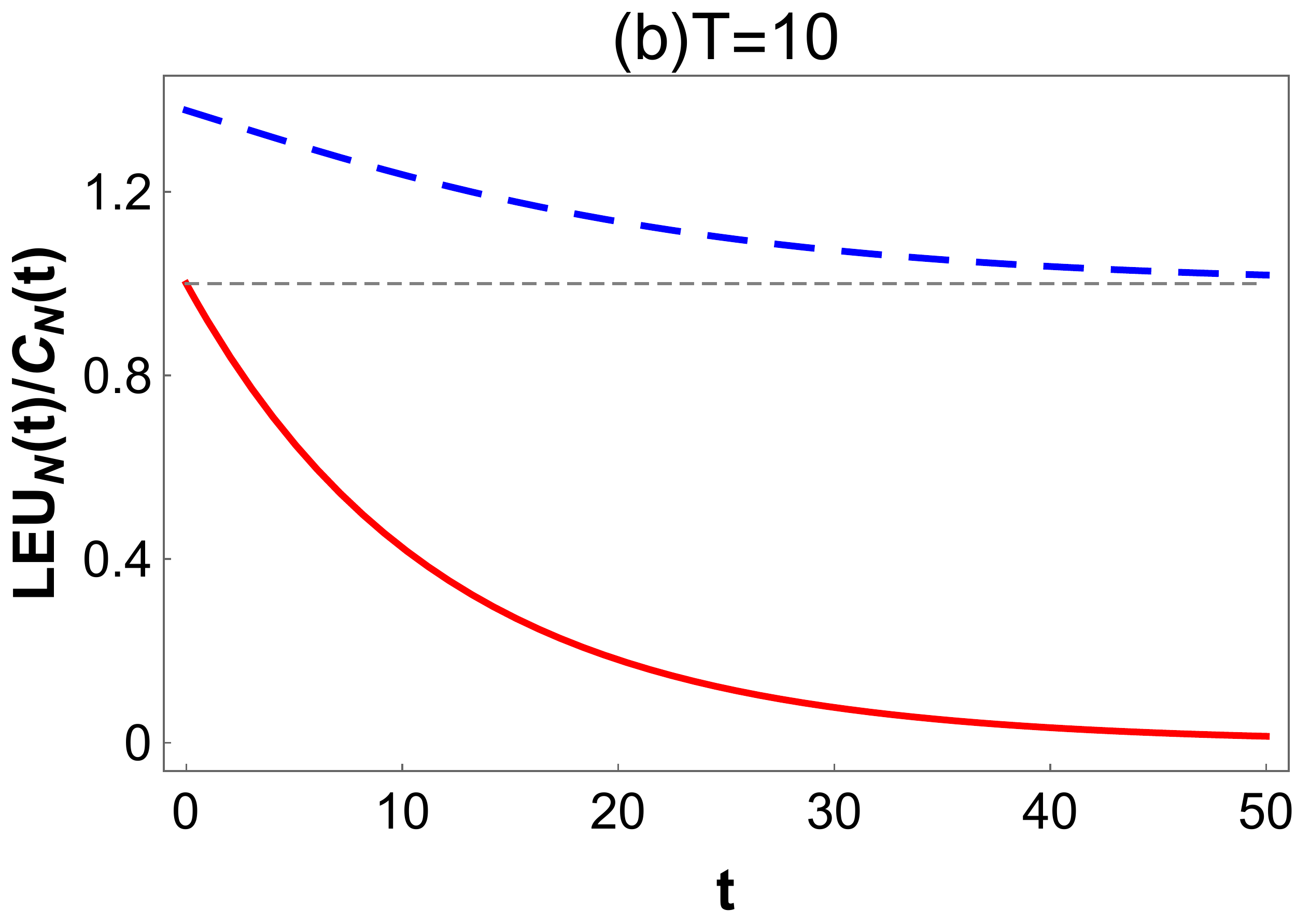}
	\centering\includegraphics[width=0.31\textwidth]{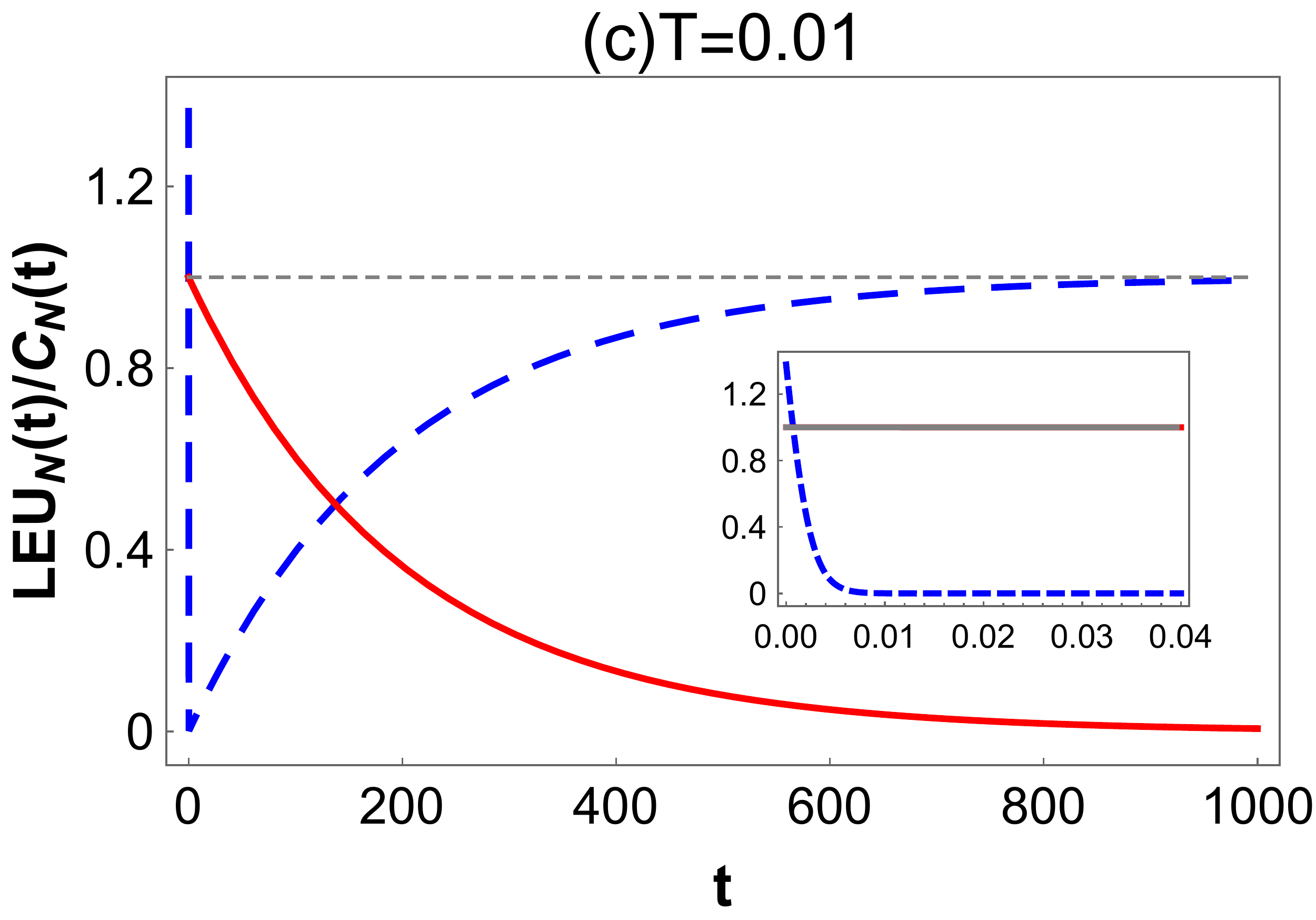}
	\caption{$(a)$ The dynamics of the $LEU(t)$(dashed) and $C(t) $(solid) in the absence of measurement; $(b)$ and $(c)$ are the dynamics of the $LER_{N}(t)$(dashed) and $C_{N}(t)$(solid) when the measuring time interval is $T=10$ and $T=0.01$, respectively. All types are at $\gamma_{1}=\gamma_{2}$.}
\end{figure*}

Fig. $3$ gives the dependence of non-Markovianity on $\lambda_{1}$ and $\lambda_{2}$. The black regime represents a Markovian quantum process, which the quantum information dissipates to environment from system. The other colors represent the non-Markovian case, where the recovery of entanglement can occur because the information can flow back from environment to system. The dividing line of different colors is a straight line, indicating that the influence of these $\gamma_{1}$ and $\gamma_{2}$ on the non-Markovianity is equivalent when $\lambda_{1}=\lambda_{2}$. We discussed three different cases earlier: $(1)\lambda_{1}=5\gamma _{1}, \lambda _{2}=5\gamma _{2}, (2)\lambda_{1}=5\gamma _{1}, \lambda _{2}=0.01\gamma _{2}, (3)\lambda_{1}=0,01\gamma _{1}, \lambda _{2}=0.01\gamma _{2}$. For $(1)$, the atom is weak coupling with two pesudomodes, the information of the atom decay quickly, thus the non-Markovianity is very small, as the black regime shown in Fig. $3$. This is the physical reason that the entanglement decreases monotonically to zero and the $LEU(t)$ monotonically increase, as the red line shown in Fig. $2(a)$.
However, for $(2)$, the atom is weak coupling with $1$-th pseudomode($\lambda_{1}=5\gamma_{1}$) and strong coupling with the $2$-th pseudomode($\lambda_{2}=0.01\gamma _{2}$), indicating that the entire environment is non-Markovian and non-Markovianity is more bigger than case $(1)$ from fig. $3$. That is why the information of atoms decay more slowly than $(1)$, as the black line shown in Fig. 1(b). For $(3)$, each atom is strong coupling with theirs two pseudomodes. Therefore, it explains that in case $(3)$, the information decay is the slowest, and the entanglement witness time is the longest, as the blue line shown in Fig. $1(b)$. That is, information can flow back from the environment to the system. In other words, in the process of strong coupling, the environment has a memory effect, which makes information flow back to the system from the environment, thus producing the phenomenon of entanglement decay and entanglement recovery. In practice, the dissipation plays a leading
role, so it decays to zero in end. It is worth noting that the distinction of strong coupling and weak coupling is different from the condition $\lambda_{1}>2\gamma_{1}, \lambda_{2}>2\gamma_{2} $, that is, $\gamma _{1}, \gamma _{2}, \lambda_{1}$, and
$\lambda_{2}$ affect the environment together, as shown in fig. $2$ and fig. $3$ for details of how they affect the whole environment.

\subsection{Entanglement Witness and Entropy Uncertainty with Zeno Effect}

For the above two-atom system, the initial state is $|\Phi \left( 0\right)
\rangle =$ $\frac{1}{\sqrt{2}}\left[ |eg\rangle +|ge\rangle \right] $. Now,
we carry out a series of measurements on the system, each measurement
interval is $T$, after $N$ measurements, the initial state survival
probability is given by%
\begin{equation}  \label{Eq29}
\begin{aligned} P_{N}\left( t\right)&=\left[ P\left( T\right) \right] ^{N}\\
&=\langle \Phi\left( 0\right) |\hat{\rho }^{\Phi }\left( T\right) |\Phi
\left(0\right) \rangle^{N} \\ &=e^{-\gamma _{z}\left( T\right) t}
\end{aligned}
\end{equation}%
where $t=NT$, and $\gamma _{z}\left( T\right) =-\frac{\log |M_{1}(T)|^{2}}{T}$ is an effective decay rate. Not only the protective
measurements affects the probability $P_{N}\left( t\right) $ but also
corrects for concurrency. And now, the effective dynamics of concurrency
depends on $T$. After performing $N$ measurements at time $t=NT$, the
concurrency is given by
\begin{equation}  \label{Eq30}
\begin{aligned}
C_{N}(t)&=|M_{1}(T)|^{2N}\\
&=e^{-\gamma _{z}\left( T\right) t}.
\end{aligned}
\end{equation}
According to the \eqref{Eq29}, $C_{N}(t)=P_{N}(t)$.

In Fig. $4$ We compare the dynamics of the $C_{N}\left( t\right) $ and $%
LEU_{N}\left( t\right) $\ with the initial maximum entangled state at
different time intervals $T$ when $\lambda _{1}=0.01\gamma
_{1}$, $\lambda _{2}=0.01\gamma _{2}$. In Fig. $4(a)$, it is absence of measurement for the dynamics of the $C(t)$ and $LEU(t)$. On the one hand, the $C(t)$ decays to zero at time approximately equal to $15$, and then there is an entanglement recovery. It is worth noting that the entanglement can be witnessed a second time after the entanglement recovery. On the other hand, the fist entangled witness time is $6.23$ and the minimal value of entanglement witness is $0.66$. For Fig. $4(b)$, i.e., $T=10$, the $C_{N}(t)$ deceases more slower than the former and it decays monotonically to zero. Nevertheless, entanglement cannot be witnessed because $LEU(t)$ is always greater than $1$. This is the anti-Zeno effect, which is normal.
For Fig. $4(c)$, i.e., $T=0.01$, the $C_{N}(t)$ is decays very slowly and $C_{N}(t)$ is greater than the previous cases. Correspondingly, the entangled witness time is around $800$, which is much greater than before. From Fig. $4(c)$, at time greater than $0$, entanglement can be witnessed. By comparing fig. $4(a)$, fig. $4(b)$ and fig. $4(c)$, it is found that the influence of Zeno effect is obviously greater than that of the environment, hence entanglement and entanglement witness can be well protected. It is known that in each case, entanglement can be effectively
protected by frequent measurements, and anti-Zeno effect will occur before
the most effective protection is achieved.

\section{CONCLUSION}
To sum up, we focus on a two-atom system under the quantum Zeno effect, in which each atom is in an environment with a double Lorentzian spectrum, and there is no interacting between the two environments. First, we use pseudomode theory to solve a model of two-level atom coupled with double Lorentzian spectrum, and obtain an analytical representation of the density operator of the atom. Secondly, we investigate the entanglement witness and entropy uncertainty, and get the analytical representation of entanglement and uncertainty and their relationship. In addition, the effects of quantum Zeno effect and environmental parameters on the entanglement witness time and the lower bound of quantum entropy uncertainty in the existence of quantum memory are discussed in detail. The results show that, only when the two spectrums satisfy strong coupling with the atom, the time of entanglement witness can be prolonged and the lower bound of the entropy uncertainty can be reduced, and the entanglement can be witnessed many times. The Zeno effect not only can very effectively prolong the time of entanglement witness and reduce the lower bound of the entropy uncertainty, but also can greatly reduce the entanglement value of witness. The results can be applied to the fields of quantum entanglement witness\cite{Koenig.2012}, quantum cryptography\cite{Jennewein},  classical correlation locking in quantum state\cite{Grosshans}and so on. We also explain the double Lorentzian with the pseudomode theory, and gave the corresponding physical explanation of total system use the non-Markovianity. 

\appendix
\section{APPENDIX: the equation in Lindblad form of an atom}

	The interaction between an atom and its environment is equivalent to the interaction between an atom and two pseudomodes. First, we construct a complete basis vector for the system comprise the an atom and two pseudomodes in the \bigskip Schr\"{o}dinger picture. The basis is

	\begin{subequations}\label{A1}
		\begin{align}
		|1\rangle&=|g,0_{1},0_{2}\rangle,\\
		|2\rangle&=|e,0_{1},0_{2}\rangle,\\
		|3\rangle&=|g,1_{1},0_{2}\rangle,\\
		|4\rangle&=|g,0_{1},1_{2}\rangle,
		\end{align}
	\end{subequations}
	where $|g\rangle(|e\rangle)$ is the basis of the atom system and $|n_{i}\rangle(n=0,1;i=1,2)$ is the introduced basis for the pseudomode $i$. For a single
	excitation of the total system, we can get their state is 
	\begin{eqnarray}\label{A2}
	|\bar{\Psi}\rangle=B_{0}(t)|1\rangle+B_{1}(t)|2\rangle+\bar{P}_{1}(t)|3\rangle+\bar{P}_{2}(t)|4\rangle.
	\end{eqnarray}
	
	On the other hand, since the number of particles is conserved. We denote the atomic ground-state population by $\bar{N}$, we have
	\begin{equation}\label{A3}
	\bar{N}=|M_{0}(t)|^2+\sum_{k}|M_{k}(t)|^2,	
	\end{equation}
	in terms of the original modes of the system. Now, we can obtain
	\begin{equation}\label{A4}
	\frac{d\bar{N}}{dt}=-\frac{d|M_{1}(t)|^2}{dt},
	\end{equation}
	which is conservation of probability. However, we now use the pseudomode Eq.\eqref{E12} to find that
	\begin{equation}\label{A5}
	\begin{aligned}
	\frac{d\bar{N}}{dt}
	=&i\bar{g}_{1}\bar{P}_{1}(t)M_{1}^{\ast}(t)-i\bar{g}_{1}\bar{P}_{1}^{\ast}(t)M_{1}(t)\\
	&+i\bar{g}_{2}\bar{P}_{2}(t)M_{1}^{\ast}(t)-i\bar{g}_{2}\bar{P}_{2}^{\ast}(t)M_{1}(t),
	\end{aligned}
	\end{equation}
	Eq.\eqref{A4} is compared with the total population growth of the pseudomode obtained from Eq.\eqref{E12}
	\begin{equation}\label{A6}
	\begin{aligned}
	\frac{d|\bar{P}_{j}|^{2}(t)}{dt}
	=&2\Im(z_{j})|\bar{P}_{j}(t)|^{2}+i\bar{g}_{j}\bar{P}_{j}(t)M_{1}^{\ast}(t)\\
	&-i\bar{g}_{j}\bar{P}_{j}^{\ast}(t)M_{1}(t),
	\end{aligned}
	\end{equation}
	so that
	\begin{equation}\label{A7}
	\begin{aligned}
	\frac{d\bar{N}}{dt}
	=&\frac{d|\bar{P}_{1}(t)|^{2}}{dt}+\frac{d|\bar{P}_{2}|^{2}(t)}{dt}\\
	&-2\Im(z_{1})|\bar{P}_{1}(t)|^{2}-2\Im(z_{2})|\bar{P}_{2}(t)|^{2}.
	\end{aligned}
	\end{equation}
	Clearly, The increase in $\bar{N}$ is not just due to the pseudomode population in pseudomode system, because the pseudomodes are lossy as shown Fig.1. 
	Therefore, the density matrix of pseudomode system is 
	\begin{equation}
	\begin{aligned}
	\bar{\rho}=&|\bar{\Psi}\rangle\langle\bar{\Psi}|+\sum_{j=1}^{2}\bar{N_{j}}|0\rangle\langle0|,\\
	\bar{N}_{j}=&2\lambda_{j}\int_{0}^{t}|\bar{P}_{j}(t^{'})|^{2}dt^{'}.
	\end{aligned}	
	\end{equation}
	
	We simply obtain the equation of Lindblad form	
	\begin{equation}\label{A9}
	\begin{aligned}
	\frac{\bar{\rho}}{dt}=&-i[H,\bar{\rho}]-\lambda_{1}[\bar{P}_{1}^{\dag}\bar{P}_{1}\bar{\rho}-2\bar{P}_{1}\bar{\rho}\bar{P}_{1}^{\dag}+\bar{\rho}\bar{P}_{1}^{\dag}\bar{P}_{1}]\\
	&-\lambda_{2}[\bar{P}_{2}^{\dag}\bar{P}_{2}\bar{\rho}-2\bar{P}_{2}\bar{\rho}\bar{P}_{2}^{\dag}+\bar{\rho}\bar{P}_{2}^{\dag}\bar{P}_{2}],
	\end{aligned}	
	\end{equation} 
	with the Hamiltonian
	\begin{equation}\label{A10}
	\begin{aligned}
	H=&\omega_{0}\sigma_{+}\sigma_{-}+\omega_{0}\bar{P}_{1}^{\dag}\bar{P}_{1}+\omega_{0}\bar{P}_{2}^{\dag}\bar{P}_{2}\\
	&+\bar{g}_{1}(\bar{P}_{1}\sigma_{+}+\sigma_{-}\bar{P}_{1}^{\dag})+\bar{g}_{2}(\bar{P}_{2}\sigma_{+}+\sigma_{-}\bar{P}_{2}^{\dag}).
	\end{aligned}	
	\end{equation}
	So we know from equation \eqref{A9} that $\lambda_{j}$ is the dissipation rate of the $j$-th pseudomode, $\bar{g}_{j}=\sqrt{\frac{\gamma _{j}\lambda _{j}}{2}}$ is the coupling strength of the atom with $j$-th pseudomode.

\section*{Disclosures}

The authors declare no conflicts of interest.


\begin{thebibliography}{10}
	\newcommand{\enquote}[1]{``#1''}
	
	\bibitem{Berta}
	M.~Berta, M.~Christandl, R.~Colbeck, J.~M. Renes, and R.~Renner, \enquote{{The
			uncertainty principle in the presence of quantum memory},}
	{\protect\JournalTitle{Nature Physics}} \textbf{6}, 659--662 (2010).
	
	\bibitem{Dupuis}
	F.~Dupuis, O.~Fawzi, and S.~Wehner, \enquote{{Entanglement Sampling and
			Applications},} {\protect\JournalTitle{IEEE Transactions on Information
			Theory}} \textbf{61}, 1093--1112 (2015).
	
	\bibitem{Koenig.2012}
	R.~Konig, S.~Wehner, and J.~Wullschleger, \enquote{{Unconditional Security From
			Noisy Quantum Storage},} {\protect\JournalTitle{IEEE Transactions on
			Information Theory}} \textbf{58}, 1962--1984 (2012).
	
	\bibitem{Jarzyna}
	M.~Jarzyna and R.~Demkowiczdobrzanski, \enquote{{True precision limits in
			quantum metrology},} {\protect\JournalTitle{New Journal of Physics}}
	\textbf{17}, 013010 (2015).
	
	\bibitem{DiVincenzo}
	D.~P. Divincenzo, M.~Horodecki, D.~Leung, J.~A. Smolin, and B.~M. Terhal,
	\enquote{{Locking Classical Correlations in Quantum States},}
	{\protect\JournalTitle{Physical Review Letters}} \textbf{92}, 067902 (2004).
	
	\bibitem{Cerf}
	N.~J. Cerf, M.~Bourennane, A.~Karlsson, and N.~Gisin, \enquote{{Security of
			quantum key distribution using d-level systems.}}
	{\protect\JournalTitle{Physical Review Letters}} \textbf{88}, 127902 (2002).
	
	\bibitem{Grosshans}
	F.~Grosshans and N.~J. Cerf, \enquote{{Continuous-variable quantum cryptography
			is secure against non-gaussian attacks},} {\protect\JournalTitle{Physical
			Review Letters}} \textbf{92}, 047905 (2004).
	
	\bibitem{Vallone}
	G.~Vallone, D.~G. Marangon, M.~Tomasin, and P.~Villoresi, \enquote{{Quantum
			randomness certified by the uncertainty principle},}
	{\protect\JournalTitle{Physical Review A}} \textbf{90}, 052327 (2014).
	
	\bibitem{Chitambar.2019}
	E.~Chitambar and G.~Gour, \enquote{{Quantum resource theories},}
	{\protect\JournalTitle{Reviews of Modern Physics}} \textbf{91}, 025001
	(2019).
	
	\bibitem{Ma.2016}
	J.~Ma, B.~Yadin, D.~Girolami, V.~Vedral, and M.~Gu, \enquote{{Converting
			Coherence to Quantum Correlations},} {\protect\JournalTitle{Physical Review
			Letters}} \textbf{116}, 160407 (2016).
	
	\bibitem{Watrous.2018}
	J.~Watrous, \emph{{The theory of quantum information}} (Cambridge University
	Press, 2018).
	
	\bibitem{Heisenberg}
	W.~Heisenberg, \enquote{{{\"{U}}ber den anschaulichen Inhalt der
			quantentheoretischen Kinematik und Mechanik},}
	{\protect\JournalTitle{Zeitschrift f{\"{u}}r Physik}} \textbf{43}, 172--198
	(1927).
	
	\bibitem{Kennard}
	E.~H. Kennard, \enquote{{Zur Quantenmechanik einfacher Bewegungstypen},}
	{\protect\JournalTitle{European Physical Journal}} \textbf{44}, 326--352
	(1927).
	
	\bibitem{Robertson}
	H.~P. Robertson, \enquote{{The Uncertainty Principle},}
	{\protect\JournalTitle{Physical Review}} \textbf{34}, 163--164 (1929).
	
	\bibitem{Deutsch}
	D.~Deutsch, \enquote{{Uncertainty in Quantum Measurements},}
	{\protect\JournalTitle{Physical Review Letters}} \textbf{50}, 631--633
	(1983).
	
	\bibitem{Kraus}
	K.~Kraus, \enquote{{Complementary observables and uncertainty relations.}}
	{\protect\JournalTitle{Physical Review D}} \textbf{35}, 3070--3075 (1987).
	
	\bibitem{Maassen1988}
	H.~Maassen and J.~Uffink, \enquote{{Generalized entropic uncertainty
			relations.}} {\protect\JournalTitle{Physical Review Letters}} \textbf{60},
	1103--1106 (1988).
	
	\bibitem{Prevedel}
	R.~Prevedel, D.~R. Hamel, R.~Colbeck, K.~A.~G. Fisher, and K.~J. Resch,
	\enquote{{Experimental investigation of the uncertainty principle in the
			presence of quantum memory and its application to witnessing entanglement},}
	{\protect\JournalTitle{Nature Physics}} \textbf{7}, 757--761 (2011).
	
	\bibitem{Li}
	C.~Li, J.~Xu, X.~Xu, K.~Li, and G.~Guo, \enquote{{Experimental investigation of
			the entanglement-assisted entropic uncertainty principle},}
	{\protect\JournalTitle{Nature Physics}} \textbf{7}, 752--756 (2011).
	
	\bibitem{Justin2014}
	J.~Dressel and F.~Nori, \enquote{{Certainty in Heisenberg's Uncertainty
			Principle: Revisiting Definitions for Estimation Errors and Disturbance},}
	{\protect\JournalTitle{Physical Review A}} \textbf{89}, 022106 (2014).
	
	\bibitem{Wehner.2008}
	S.~Wehner, C.~Schaffner, and B.~M. Terhal, \enquote{{Cryptography from Noisy
			Storage},} {\protect\JournalTitle{Physical Review Letters}} \textbf{100},
	220502 (2008).
	
	\bibitem{Guhne.2009}
	O.~G{\"{u}}hne and G.~T{\'{o}}th, \enquote{{Entanglement detection},}
	{\protect\JournalTitle{Physics Reports}} \textbf{474}, 1--75 (2009).
	
	\bibitem{F.B.M.DosSantos.2013}
	F.~B. M.~D. Santos and A.~M.~S. Macedo, \enquote{{Tuning Entanglement Patterns
			in Qubits Clusters},} {\protect\JournalTitle{Journal of Quantum Information
			Science}} \textbf{2013}, 85--92 (2013).
	
	\bibitem{Amico}
	L.~Amico, R.~Fazio, A.~Osterloh, and V.~Vedral, \enquote{{Entanglement in
			many-body systems},} {\protect\JournalTitle{Reviews of Modern Physics}}
	\textbf{80}, 517--576 (2008).
	
	\bibitem{Bouwmeester.1997}
	D.~Bouwmeester, J.~Pan, K.~Mattle, M.~Eibl, H.~Weinfurter, and A.~Zeilinger,
	\enquote{{Experimental quantum teleportation},}
	{\protect\JournalTitle{Nature}} \textbf{390}, 575--579 (1997).
	
	\bibitem{Bouwmeester.2000}
	D.~Bouwmeester and A.~Zeilinger, \enquote{{The physics of quantum information:
			basic concepts},} in \emph{The physics of quantum information,}  (Springer,
	2000), pp. 1--14.
	
	\bibitem{Jennewein}
	T.~Jennewein, C.~Simon, G.~Weihs, H.~Weinfurter, and A.~Zeilinger,
	\enquote{{Quantum Cryptography with Entangled Photons},}
	{\protect\JournalTitle{Physical Review Letters}} \textbf{84}, 4729--4732
	(2000).
	
	\bibitem{Jozsa}
	A.~S{\o}rensen and K.~M{\o}lmer, \enquote{{Entanglement and quantum computation
			with ions in thermal motion},} {\protect\JournalTitle{Physical Review A}}
	\textbf{62}, 022311 (2000).
	
	\bibitem{Adam2015}
	A.~Miranowicz, K.~Bartkiewicz, N.~Lambert, Y.~Chen, and F.~Nori,
	\enquote{{Increasing relative nonclassicality quantified by standard
			entanglement potentials by dissipation and unbalanced beam splitting},}
	{\protect\JournalTitle{Physical Review A}} \textbf{92}, 062314 (2015).
	
	\bibitem{Yueh2013}
	Y.-N. Chen, S.-L. Chen, N.~Lambert, C.-M. Li, G.-Y. Chen, and F.~Nori,
	\enquote{{Entanglement swapping and testing quantum steering into the past
			via collective decay},} {\protect\JournalTitle{Physical Review A}}
	\textbf{88}, 052320 (2013).
	
	\bibitem{Ye2019}
	Y.~Chen, W.~Qin, and F.~Nori, \enquote{{Fast and high-fidelity generation of
			steady-state entanglement using pulse modulation and parametric
			amplification},} {\protect\JournalTitle{Physical Review A}} \textbf{100},
	012329 (2019).
	
	\bibitem{Wei2018}
	W.~Qin, A.~Miranowicz, P.~Li, X.~Lu, J.~Q. You, and F.~Nori,
	\enquote{{Exponentially Enhanced Light-Matter Interaction, Cooperativities,
			and Steady-State Entanglement Using Parametric Amplification},}
	{\protect\JournalTitle{Physical Review Letters}} \textbf{120}, 093601 (2018).
	
	\bibitem{Breuer}
	H.-P. Breuer and F.~Petruccione, \emph{{The theory of open quantum systems}}
	(Oxford University Press on Demand, 2002).
	
	\bibitem{Duan.2000}
	L.~M. Duan, A.~S. Sorensen, J.~I. Cirac, and P.~Zoller, \enquote{{Squeezing and
			entanglement of atomic beams.}} {\protect\JournalTitle{Physical Review
			Letters}} \textbf{85}, 3991--3994 (2000).
	
	\bibitem{Weiss}
	S.~Dattagupta and S.~Puri, \enquote{{Dissipative Phenomena in Condensed
			Matter},} in \emph{Quantum Dissipative Systems,}  (Springer Berlin
	Heidelberg, 2004), pp. 173--203.
	
	\bibitem{Yu}
	T.~Yu and J.~H. Eberly, \enquote{{Quantum open system theory: bipartite
			aspects.}} {\protect\JournalTitle{Physical Review Letters}} \textbf{97},
	140403 (2006).
	
	\bibitem{Hu}
	M.~Hu and H.~Fan, \enquote{{Quantum-memory-assisted entropic uncertainty
			principle, teleportation, and entanglement witness in structured
			reservoirs},} {\protect\JournalTitle{Physical Review A}} \textbf{86}, 032338
	(2012).
	
	\bibitem{Zou}
	H.~Zou, M.~Fang, B.~Yang, Y.~Guo, W.~He, and S.~Zhang, \enquote{{The quantum
			entropic uncertainty relation and entanglement witness in the two-atom system
			coupling with the non-Markovian environments},}
	{\protect\JournalTitle{Physica Scripta}} \textbf{89}, 115101 (2014).
	
	\bibitem{Francica.2010}
	F.~Francica, F.~Plastina, and S.~Maniscalco, \enquote{{Quantum Zeno and
			anti-Zeno effects on quantum and classical correlations},}
	{\protect\JournalTitle{Physical Review A}} \textbf{82}, 052118 (2010).
	
	\bibitem{Maniscalco.2008}
	S.~Maniscalco, F.~Francica, R.~L. Zaffino, N.~L. Gullo, and F.~Plastina,
	\enquote{{Protecting Entanglement via the Quantum Zeno Effect},}
	{\protect\JournalTitle{Physical Review Letters}} \textbf{100}, 090503 (2008).
	
	\bibitem{Petrosky1991}
	T.~Petrosky, S.~Tasaki, and I.~Prigogine, \enquote{{Quantum Zeno effect},}
	{\protect\JournalTitle{Physica A-statistical Mechanics and Its Applications}}
	\textbf{170}, 306--325 (1991).
	
	\bibitem{Qing2013}
	Q.~Ai, D.~Xu, S.~Yi, A.~G. Kofman, C.~P. Sun, and F.~Nori, \enquote{{Quantum
			anti-Zeno effect without wave function reduction},}
	{\protect\JournalTitle{Scientific Reports}} \textbf{3}, 01752 (2013).
	
	\bibitem{Lan2009}
	L.~Zhou, S.~Yang, Y.~Liu, C.~P. Sun, and F.~Nori, \enquote{{Quantum Zeno switch
			for single-photon coherent transport},} {\protect\JournalTitle{Physical
			Review A}} \textbf{80}, 062109 (2009).
	
	\bibitem{Behzadi.2017}
	N.~Behzadi, B.~Ahansaz, A.~Ektesabi, and E.~Faizi, \enquote{{Controlling
			speedup in open quantum systems through manipulation of system-reservoir
			bound states},} {\protect\JournalTitle{Physical Review A}} \textbf{95},
	052121 (2017).
	
	\bibitem{Fanchini.2013}
	F.~F. Fanchini, G.~Karpat, L.~K. Castelano, and D.~Z. Rossatto,
	\enquote{{Probing the degree of non-Markovianity for independent and common
			environments},} {\protect\JournalTitle{Physical Review A}} \textbf{88},
	012105 (2013).
	
	\bibitem{Calajo.2017}
	G.~Calajo, L.~Rizzuto, and R.~Passante, \enquote{{Control of spontaneous
			emission of a single quantum emitter through a time-modulated
			photonic-band-gap environment},} {\protect\JournalTitle{Physical Review A}}
	\textbf{96}, 023802 (2017).
	
	\bibitem{Garraway.1997}
	B.~M. Garraway, \enquote{{Nonperturbatiive decay of an atomic system in a
			cavity},} {\protect\JournalTitle{Physical Review A}} \textbf{55}, 013010
	(1997).
	
	\bibitem{Garraway.1996}
	B.~M. Garraway and P.~L. Knight, \enquote{{Cavity modified quantum beats},}
	{\protect\JournalTitle{Physical Review A}} \textbf{54}, 3592--3602 (1996).
	
	\bibitem{Mazzola.2009}
	L.~Mazzola, S.~Maniscalco, J.~Piilo, K.~Suominen, and B.~M. Garraway,
	\enquote{{Pseudomodes as an effective description of memory: Non-Markovian
			dynamics of two-state systems in structured reservoirs},}
	{\protect\JournalTitle{Physical Review A}} \textbf{80}, 012104 (2009).
	
	\bibitem{Bellomo}
	B.~Bellomo, R.~L. Franco, and G.~Compagno, \enquote{{Entanglement dynamics of
			two independent qubits in environments with and without memory},}
	{\protect\JournalTitle{Physical Review A}} \textbf{77}, 032342 (2008).
	
	\bibitem{Bose.2001}
	S.~Bose, I.~Fuentes, P.~Knight, and V.~Vedral, \enquote{{Erratum: Subsystem
			Purity as an Enforcer of Entanglement [Phys. Rev. Lett. 87, 050401 (2001)]},}
	{\protect\JournalTitle{Physical Review Letters}} \textbf{87}, 279901 (2001).
	
	\bibitem{Pratt2004}
	J.~S. Pratt, \enquote{{Universality in the Entanglement Structure of
			Ferromagnets},} {\protect\JournalTitle{Physical Review Letters}} \textbf{93},
	237205 (2004).
	
	\bibitem{Hagley.1997}
	E.~Hagley, X.~Maitre, G.~Nogues, C.~Wunderlich, M.~Brune, J.~M. Raimond, and
	S.~Haroche, \enquote{{Generation of einstein-podolsky-rosen pairs of atoms},}
	{\protect\JournalTitle{Physical Review Letters}} \textbf{79}, 1--5 (1997).
	
	\bibitem{Wootters.1998}
	W.~K. Wootters, \enquote{{Entanglement of formation of an arbitrary state of
			two qubits},} {\protect\JournalTitle{Physical Review Letters}} \textbf{80},
	2245--2248 (1998).
	
	\bibitem{Koshino.2005}
	K.~Koshino and A.~Shimizu, \enquote{{Quantum Zeno effect by general
			measurements},} {\protect\JournalTitle{Physics Reports}} \textbf{412},
	191--275 (2005).
	
	\bibitem{Facchi.2001}
	P.~Facchi, H.~Nakazato, and S.~Pascazio, \enquote{{From the quantum zeno to the
			inverse quantum zeno effect.}} {\protect\JournalTitle{Physical Review
			Letters}} \textbf{86}, 2699--2703 (2001).
	
	\bibitem{Maniscalco.2006}
	S.~Maniscalco, J.~Piilo, and K.~Suominen, \enquote{{Zeno and anti-Zeno effects
			for quantum Brownian motion.}} {\protect\JournalTitle{Physical Review
			Letters}} \textbf{97}, 130402 (2006).
	
	\bibitem{Bernu.2008}
	J.~Bernu, S.~Deleglise, C.~Sayrin, S.~Kuhr, I.~Dotsenko, M.~Brune, J.~M.
	Raimond, and S.~Haroche, \enquote{{Freezing coherent field growth in a cavity
			by the quantum zeno effect},} {\protect\JournalTitle{Physical Review
			Letters}} \textbf{101}, 180402 (2008).
	
	\bibitem{Kang2019}
	K.~Wu, Z.~Hou, G.~Xiang, C.~Li, G.~Guo, D.~Dong, and F.~Nori,
	\enquote{{Detecting Non-Markovianity via Quantified Coherence: Theory and
			Experiments},} {\protect\JournalTitle{npj Quantum Inf}}  (2019).
	
	\bibitem{Wei2012}
	W.-m. Zhang, P.-y. Lo, H.-n. Xiong, M.~W.-y. Tu, and F.~Nori, \enquote{{General
			Non-Markovian Dynamics of Open Quantum Systems},}
	{\protect\JournalTitle{Physical Review Letters}} \textbf{109}, 170402 (2012).
	
	\bibitem{Breuer2010}
	E.~Laine, J.~Piilo, and H.~Breuer, \enquote{{Measure for the Non-Markovianity
			of Quantum Processes},} {\protect\JournalTitle{Physical Review A}}
	\textbf{81}, 062115 (2010).
	
	\bibitem{He.2017}
	Z.~He, H.~Zeng, Y.~Li, Q.~Wang, and C.~Yao, \enquote{{Non-Markovianity measure
			based on the relative entropy of coherence in an extended space},}
	{\protect\JournalTitle{Physical Review A}} \textbf{96}, 022106 (2017).
	
\end{thebibliography}
\end{document}